\documentclass[journal,twoside, web]{ieeecolor}

\usepackage[table,xcdraw]{xcolor}
\usepackage{cite}
\usepackage{generic}
\usepackage{amsmath,amssymb,amsfonts}
\usepackage{graphicx}
\usepackage{stfloats}
\usepackage{hyperref}
\hypersetup{
    colorlinks=true,
    linkcolor=blue,
    filecolor=magenta,      
    urlcolor=cyan,
}

\graphicspath{{Figures/}}
\newcommand{\com}[1]{``#1''}

\begin{document}
\title{L-SFAN: Lightweight Spatially-focused Attention Network for Pain Behavior Detection}
\author{Jorge~Ortigoso-Narro, Fernando~Diaz-de-Maria~(\IEEEmembership{Member, IEEE}), Mohammad Mahdi~Dehshibi~(\IEEEmembership{Senior Member, IEEE}) and Ana Tajadura-Jim\'{e}nez
\thanks{This research has been funded through various sources. The Spanish Ministry of Science and Innovation (State Research Agency) provided partial funding through National Grant PID2020-118504GB-I00. Madrid Regional Government also provided partial funding through grants Y2020/NMT-6660 for the interdisciplinary project COMPANION-CM and REACT UE Grant IntCARE-CM for the Intelligent and Interactive Home Care System to mitigate the COVID-19 pandemic. The European Research Council (ERC) also provided funding through the European Union’s Horizon 2020 research and innovation program with grant agreement No 101002711 for the BODYinTRANSIT project.}
\thanks{Jorge Ortigoso-Narro is affiliated with the Department of Signal Theory and Communications, Universidad Carlos III de Madrid in Legan\'{e}s, Spain (e-mail: jortigoso@tsc.uc3m.es). }
\thanks{Fernando Diaz-de-Maria is affiliated with the Department of Signal Theory and Communications, Universidad Carlos III de Madrid in Legan\'{e}s, Spain (e-mail: fdiaz@ing.uc3m.es).}
\thanks{Mohammad Mahdi Dehshibi is affiliated with the Department of Computer Science and Engineering, Universidad Carlos III de Madrid in Legan\'{e}s, Spain, as well as the Unconventional Computing Laboratory at the University of the West of England in Bristol, U.K. (e-mail: mohammad.dehshibi@yahoo.com).}
\thanks{Ana Tajadura-Jim\'{e}nez is affiliated with 
the Department of Computer Science and Engineering, Universidad Carlos III de Madrid in Legan\'{e}s, Spain, as well as the UCL Interaction Centre, University College London in London, U.K. (e-mail: atajadur@inf.uc3m.es).}}

\maketitle

\begin{abstract}
Chronic Low Back Pain (CLBP) afflicts millions globally, significantly impacting individuals' well-being and imposing economic burdens on healthcare systems. While artificial intelligence (AI) and deep learning offer promising avenues for analyzing pain-related behaviors to improve rehabilitation strategies, current models, including convolutional neural networks (CNNs), recurrent neural networks, and graph-based neural networks,  have limitations. These approaches often focus singularly on the temporal dimension or require complex architectures to exploit spatial interrelationships within multivariate time series data. To address these limitations, we introduce \hbox{L-SFAN}, a lightweight CNN architecture incorporating 2D filters designed to meticulously capture the spatial-temporal interplay of data from motion capture and surface electromyography sensors. Our proposed model, enhanced with an oriented global pooling layer and multi-head self-attention mechanism, prioritizes critical features to better understand CLBP and achieves competitive classification accuracy. Experimental results on the EmoPain database demonstrate that our approach not only enhances performance metrics with significantly fewer parameters but also promotes model interpretability, offering valuable insights for clinicians in managing CLBP. This advancement underscores the potential of AI in transforming healthcare practices for chronic conditions like CLBP, providing a sophisticated framework for the nuanced analysis of complex biomedical data.
\end{abstract}


\section{Introduction}
\label{sec:introduction}

\IEEEPARstart{C}{hronic} pain (CP), particularly chronic low back pain (CLBP), represents a growing concern worldwide, severely affecting individuals' quality of life by limiting professional capabilities, physical activities, and psychological well-being. With 20\% of adults in the United States suffering CP, the condition not only affects individuals but also imposes significant socio-economic impacts~\cite{Yong2021, ROY2022123, Fayaz2016}. Physical rehabilitation, a pivotal aspect of CP management, traditionally relies on self-management programs under the periodic supervision of medical professionals, such as physiotherapists~\cite{Aneesha}. However, access to effective rehabilitation in traditional medical settings is significantly limited, compounding the challenges CP patients face. These individuals often alter their movements and activities to avoid pain, leading to behaviors that exacerbate disability and diminish the quality of life~\cite{Leeuw}. This adaptation, particularly prevalent in CLBP patients~\cite{KEEFE1982363}, underscores the critical need for adaptable therapeutic strategies that can be effectively implemented beyond the confines of conventional care environments~\cite{VLAEYEN2000317}.

Recent advancements in artificial intelligence (AI) and deep learning~\cite{ashtari2022multi, dehshibi2023deep} can transcend traditional care settings, enabling remote and personalized therapeutic interventions. These technologies offer promising opportunities for developing diagnostic tools to help clinicians enhance the personalization and effectiveness of rehabilitation strategies. However, existing models, such as Stacked-LSTMs~\cite{lstm2}, GRU-based sparsely-connected recurrent-based neural networks~\cite{moha}, graph-based neural networks~\cite{wang2021leveraging}, and spatio-temporal attention networks~\cite{wang2019learning}, frequently struggle to focus on the most relevant spatiotemporal patterns in pain-related behaviors. This limitation can be attributed to their reliance on architectures that are either not properly designed to effectively find those relevant hidden patterns or require large computational resources, often resulting in inefficient and unsustainable models with significant computational and energy costs~\cite{Martineau2020}. On the other hand, approaches that use multimodal information and feature engineering~\cite{emopainmultimodalmultilevel} or multilevel context~\cite{emopainmultilevelcontext} do not effectively prioritize the data patterns most critical for detecting pain-related behaviors.

To overcome these limitations, we propose an end-to-end lightweight CNN architecture with 2D convolutional filters to analyze the nuanced spatiotemporal interactions captured by motion capture (MoCAP) and surface electromyography sensors (sEMG) sensors in CLBP patients. By spatial patterns, we refer to the relationships between the outputs of the sensors arranged over the patient’s body, and by temporal patterns, we refer to the temporal evolution of these outputs. This architecture, enhanced with an oriented global pooling layer and a multi-head self-attention mechanism, prioritizes essential features to better understand pain-related behaviors and ensures model efficiency. We use Grad-CAM~\cite{Selvaraju_2017_ICCV} to elucidate the spatial patterns critical to the diagnosis of CLBP and provide deeper insights into the model’s decision-making processes to improve the clinical utility of our model. Finally, by demonstrating our model’s effectiveness on the EmoPain database~\cite{7173007}, we provide an open benchmark for leveraging AI to address chronic conditions like CLBP.

Our research particularly focuses on protective behavior detection (PBD) and expands the boundaries of conventional approaches in several significant ways:

\begin{enumerate}
  \item \textbf{Lightweight Sustainable Architecture:} We have developed a lightweight 2D-CNN architecture that can efficiently analyze CLBP through motion capture and sEMG data. This approach advances the field technologically and provides an effective and scalable solution with the potential for real-world clinical applications.
  \item \textbf{Spatial Pattern Analysis for detection of pain-related behaviors:} Our research highlights the critical role of spatial patterns in detecting pain-related behaviors. Our model enhances the understanding of the manifestation of protective behaviors in movement, which can help refine diagnostic criteria and develop targeted therapeutic strategies.
  \item \textbf{Benchmark Performance with Lightweight and Scalable Model:} Our model achieves competitive performance on the EmoPain dataset while being much simpler and more efficient than current state-of-the-art systems. Our research contributes to making advanced AI diagnostics accessible in resource-constrained settings.
\end{enumerate}

The rest of this paper is organized as follows: Section~\ref{sec:related_work} reviews previous studies using the EmoPain database. Section~\ref{sec:proposed_system} elaborates on the proposed method in depth. Section~\ref{sec:experiments_results} presents and discusses the experimental setup and results. Finally, Section~\ref{sec:coclusion} concludes the paper and suggests directions for future research.

\section{Related work}
\label{sec:related_work}
Initial research using the EmoPain dataset~\cite{7173007} pointed towards the utility of RNN architectures in detecting pain levels and recognizing protective behaviors, a premise further expanded by other researchers. Through a series of studies, Wang et al. transitioned from basic LSTM models~\cite{wang2019learning} to more complex structures such as Dual-Stream LSTMs~\cite{lstm2} and Stacked-LSTMs~\cite{Wang_2021} for processing of body movement data along with data augmentation and segmentation window width approaches. This progression demonstrated the versatility of LSTMs and their ability to improve pain behavior detection through refined temporal and multimodal processing. In~\cite{lstm2}, two sets of three LSTM layers were used to analyze the MoCap and sEMG data. Experiments demonstrated the potential for better architecture to accommodate different data types.
Additionally, the impact of sliding window lengths on detection performance is discussed, with evidence that the choice should be based on knowledge of the dataset, as it is affected by the duration and complexity of the movement. In~\cite{Wang_2021}, recurrent networks with three LSTM units were explored. The experimental results revealed that the stacked-LSTM outperforms the dual-stream LSTM and CNN-based models.

The development of BodyAttentionNet (BANet)~\cite{wang2019learning} highlighted the importance of focusing on a subset of joint angles and using bodily attention mechanisms to capture the most informative temporal and body configurational cues characterizing specific movements and strategies. The proposed architecture demonstrates a substantial improvement in PBD performance and a significant reduction in the number of parameters compared to other LSTM-based architectures. The integration of Human Activity Recognition (HAR) with protective behavior detection for continuous data in chronic pain management was proposed in~\cite{wang2021leveraging}. The authors suggested a hierarchical HAR-PBD architecture using a graph convolution network and an LSTM with a class-balanced focal categorical cross-entropy loss to alleviate class imbalances. This study highlighted the potential use cases of the proposed architecture in chronic pain management and discussed the limitations of current methods that rely on pre-segmented activities.

Li et al.~\cite{PLAAN} introduced a novel anomaly detection-based system using an LSTM deep neural network with LSTM units to automatically assess chronic pain intensity and PBD from body movements within the sparse occurrences in the dataset. This method addresses the limitation of imbalanced expert-labeled data for pain estimation and proposes an anomaly detection-based approach to enhance the network's performance. The study also delved into joint training with exercise type labels, data balancing techniques, hierarchical classification for pain intensity estimation, and the cascaded estimation of protective behavior and pain intensity. Dehshibi et al.~\cite{moha} discussed the challenges of analyzing biometric data, specifically in individuals with chronic pain, and proposed a method to classify pain levels and pain-related behavior. The proposed method involves a sparsely connected recurrent neural network ensemble with gated recurrent units and incorporates information-theoretic features to compensate for variations in the temporal dimension. 

The exploration into multi-level fusion approaches by Uddin and Canavan~ \cite{emopainmultimodalmultilevel} and Phan et al. ~\cite{emopainmultilevelcontext} introduced a nuanced understanding of pain behavior prediction. These authors presented an approach for pain estimation and PBD using a multimodal, multi-level fusion of movement data. 

\begin{figure*}[!htbp]
  \centering
  {\includegraphics[width=0.9\linewidth,scale=0.37]{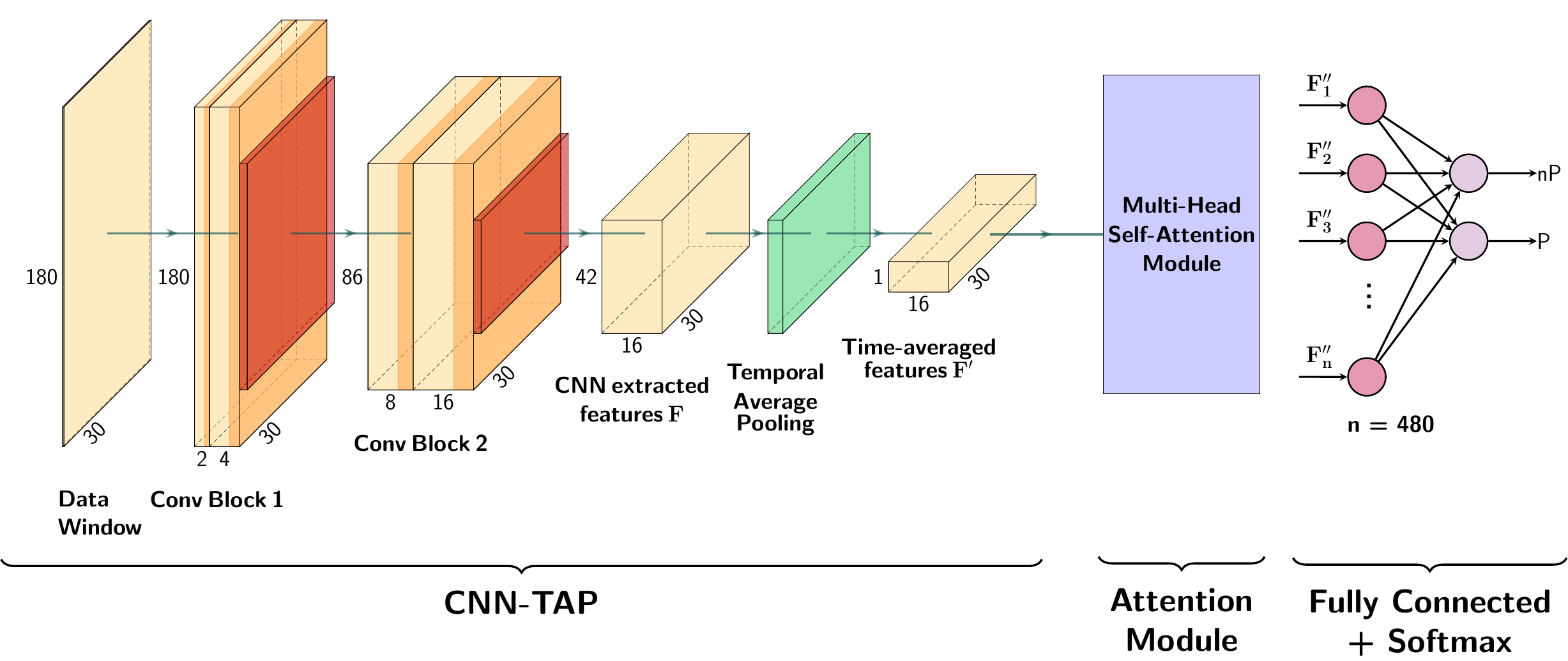}}
  \caption{The schematic of the proposed L-SFAN architecture for protective behavior detection. The $180\times30$ input matrix (13 \textbf{Joint Angles} + 13 \textbf{Joint Energies} from MoCAP IMUs + 4 sEMG outputs) is processed by a CNN-TAP backbone for feature extraction. A multi-head self-attention module further refines the features, which feed into a linear layer with softmax, providing probabilities for protective behavior (P) and its complement (nP).}
  \label{fig:architecture}
\end{figure*}

Phan et al. \cite{emopainmultimodalmultilevel} proposed a deep learning approach based on CNN and LSTM networks to extract multi-level context information from physiological signals to distinguish between pain and painlessness activities. A 1D CNN was used to extract spatial information where a pooling layer was responsible for dimension reduction. The extracted spatial information was then fed to a bidirectional LSTM network to process the temporal dimension of the data. After extracting the context vector, they used variant attention~\cite{bahdanau2016neural} after the spatio-temporal processing. Experimental results demonstrated that multi-level context information performs significantly better than uni-level context information for comprehensively analyzing spatio-temporal physiological signals.

The inclusion of contextual information for personalized assessment was explored in~\cite{10182360}, where the researchers leveraged information about the patient's current health condition to predict the pain level assessment. While this method explores the integration of a self-assessment framework, we find that its reliance on the patient's state beforehand makes it unsuitable for direct comparison with our work.

\section{Proposed Architecture}
\label{sec:proposed_system}

In this study, we adopt a data preparation process proposed by Wang et al. ~\cite{Wang_2021}. We use a sliding window technique on continuous recordings to extract 3-second sEMG and MoCAP data intervals. These intervals, captured as $180 \times 30$ matrices, reflect a detailed compilation of 60 frames per second over 3 seconds, including sEMG readings, \textbf{Joint Angles}, and \textbf{Joint Energies} derived from anatomical positions of IMU sensors~\cite{7173007}. We also strategically omitted activity type considerations to ensure robust generalization across varied trials, focusing on detecting protective behaviors within each data window. Section~\ref{sec:experiments_results} further discusses the data preprocessing and representation.

The proposed architecture, named Lightweight Spatially-Focused Attention Network (L-SFAN), comprised two main parts: a 2D CNN feature extractor backbone and an attention mechanism to further refine the features for protective behavior detection. Fig.~\ref{fig:architecture} depicts the proposed architecture, including details regarding the number of layers and dimensions of the activation maps.

\subsection{L-SFAN -- Feature Extraction Module}
In this study, we propose an end-to-end lightweight 2D-CNN architecture tailored to the specific challenges of multivariate time series classification. While CNNs with 2D filters are usually applied to computer vision tasks~\cite{alexnet,long2014fully,dehshibi2024beenet,dehshibi2023ADVISE, ashtarimajlan2023deep}, the ability of 2D-CNNs to identify patterns within two-dimensional data extends to other domains. In particular, they have shown promising results in capturing the intricate relationships between multivariate time series~\cite{7870510}. The \hbox{L-SFAN} architecture is designed to exploit the complex relationships between sEMG and bodily-arranged motion sensor modalities and their temporal dynamics. 

We have configured our sensory data to form a 2D matrix of size $180 \times 30$. This specific data configuration incorporates temporal dynamics (i.e., 3 seconds of data at 60 Hz) and spatial information, including IMU and sEMG sensor locations and arrangements (i.e., Joint Angles and Joint Energies). For clarity, let us denote our time window of data as $\mathbf{X} = [s^{(k)}_{1}, s^{(k)}_{2}, \cdots, s^{(k)}_{L}]$, where $s^{(k)}_{t}$ represents a $k$-dimensional feature vectors at time step $t$ over a time span $0<t<L$ with a window length of $L$.

\begin{figure}[!htbp]
  \centering
	{\includegraphics[width=1\linewidth]{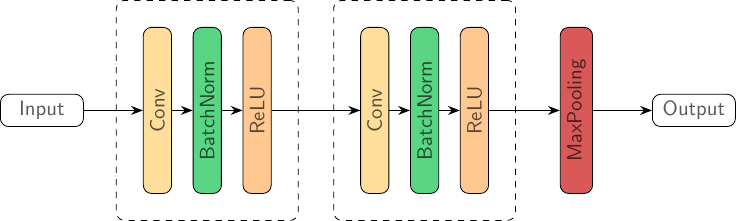}}
  \caption{The convolutional block used in the feature extractor module.}
  \label{fig:convblock}
\end{figure}

The core strength of the L-SFAN lies in its lightweight feature extractor backbone, which comprises two basic convolutional blocks and learns hierarchical feature representations through convolutional and pooling layers. Each convolutional block comprises a sequence of two convolutional layers, each followed by a batch normalization layer and the corresponding activation function, and concludes with a max-pooling layer (see Fig.~\ref{fig:convblock}).

Convolutional layers employ a set of $C$ filters (i.e., kernels) to extract distinctive local features or patterns from the input data and produce corresponding feature maps, capturing the spatial distribution of the detected features. Mathematically, the operation of a convolutional layer is expressed in Eq.~\ref{eq:convlayer}.

\begin{equation}
  \label{eq:convlayer}
  Y^{(c)}_{i,j} = f \left( \sum_{m,n} X_{(i+m,j+n)} \cdot W^{(c)}_{m,n} + b^{(c)} \right)
\end{equation}
where $Y^{(c)}_{i,j}$ represents the feature map at spatial position $(i,j)$ resulting from applying the $c^{th}$ kernel, $X_{(i+m,j+n)}$ denotes the input data values within the kernel region, $W^{(c)}$ represents the weights associated with the $c^{th}$ kernel, $b^{(c)}$ is the bias term for the $c^{th}$ kernel, and $f$ denotes a non-linear activation function.

The activation function used in this work is the Rectified Linear Unit (ReLU)~\cite{loffe2015} defined as $\mathrm{ReLU}(x) = \max(0,x)$)). Following this, a pooling layer is incorporated to reduce the feature map dimensionality, increase the receptive field, and enhance robustness against minor input variations. 

We use a kernel size of $(3, 3)$ with zero-padding to maintain the spatial dimension of the feature maps. The zero-padding ensures the preservation of all 30 dimensions along the sensor (spatial) axis, which is crucial for retaining full resolution in the sEMG and joint data. To solely reduce the temporal dimension, we employ a max-pooling layer with a kernel size of $(2,1)$ and stride of $(2,1)$. Consequently, the feature extractor backbone transforms the input matrix $\mathbf{X} \in \mathbb{R}^{180\times30}$ into a feature tensor $\mathbf{F} \in \mathbb{R}^{16\times42\times30}$, where $16$ represents the number of feature maps, $42$ the temporal resolution, and $30$ the spatial resolution.

In this work, we aim to capitalize on the potential benefits of preserving spatial resolution while analyzing temporal sensorial data. To achieve this, we propose using \textbf{Temporal Average Pooling (TAP)}. The pooling strategy, illustrated as CNN-TAP in Fig.~\ref{fig:architecture}, involves averaging the feature tensor $\mathbf{F}$ across the temporal dimension. This results in $\mathbf{F'} \in \mathbb{R}^{16 \times 30}$, which retains the spatial resolution, i.e., the sensor dimension. This preservation of spatial information is crucial for identifying which sensor inputs contribute most significantly to PBD.
\begin{figure*}[!b]
  \centering
    {\includegraphics[width=0.9\linewidth,scale=0.65]{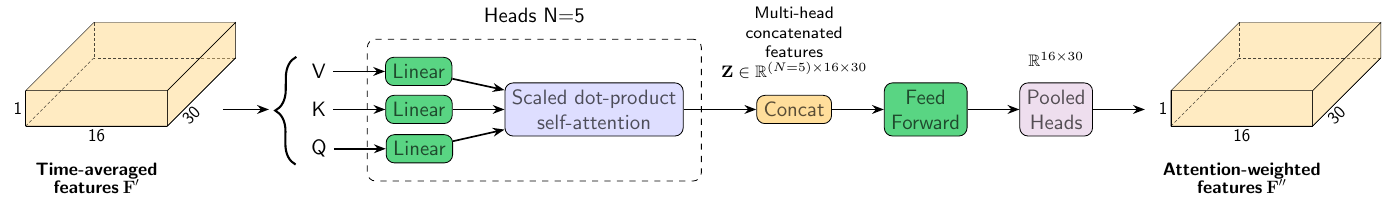} }
  \caption{Block diagram of the proposed multi-head attention module.}
  \label{fig:multihead}
\end{figure*}

\subsection{L-SFAN -- Self-Attention Module}
The self-attention mechanism empowers deep learning models to effectively handle sequential data by selectively focusing on relevant input regions based on their similarity to other parts. Prior research has also demonstrated the benefits of incorporating attention mechanisms for protective behavior detection~\cite{wang2019learning, wang2021leveraging}. In our case, we propose using self-attention to effectively capture the relevant interdependencies among the spatially-focused features extracted by our CNN-TAP backbone.

Formally, we calculate Queries ($Q$), Keys ($K$) and Values ($V$) with respect to the input sequence $X$ using Eq.~\ref{eq:att_eqs}, where $W_q$, $W_k$ and $W_v$ represent learnable weight matrices.

\begin{equation}
  \label{eq:att_eqs}
  \begin{aligned}
    Q &= W_q\cdot X \\ K &= W_k\cdot X \\ V &= W_v\cdot X
  \end{aligned}
\end{equation}

Having calculated the query vector $Q$, we compute the affinity score. This score represents the similarity between the query and key, indicating the level of attention required for the corresponding value. The affinity score $A$ for each input position is calculated using Eq.~\ref{eq:att_score}.

\begin{equation}
    \label{eq:att_score}
    A = \frac{Q\cdot K^{T}}{\sqrt{d_{k}}}
\end{equation}
where $\sqrt{d_{k}}$ serves as a normalization factor that ensures the dot product result, $Q\cdot K^{T}$, is independent of the temporal dimension length ($d_{k} = 180$ in this study). Subsequently, a \textit{softmax} function is applied to obtain normalized attention weights that sum to 1 for each query. Finally, the self-attention output is computed as a weighted sum of the value vectors, as described in Eq.~\ref{eq:self_att}.

\begin{equation}
    \label{eq:self_att}
    Z = softmax\left(A\right)\cdot V
\end{equation}

In the multi-head setting, this operation is repeated with different learned weight matrices ($W_q^{(i)}$, $W_k^{(i)}$, $W_v^{(i)}$) for each head $i=1, \cdots, N$. The resulting multiple outputs are then concatenated and transformed linearly using a feedforward network.

Given the pre-existing informative representation generated by the CNN-TAP feature extraction backbone, we opted not to use an additional embedding layer before the attention module. Instead, we used five heads ($N=5$), a selection supported by preliminary experiments and previous research suggesting their effectiveness and efficiency~\cite{9310460, LI202114}. These heads feed a feedforward layer of dimension 30 with a dropout probability 0.2. This process yields an enhanced feature tensor, $\mathbf{F''} \in \mathbb{R}^{16\times30}$, that leverages the self-similarities within our prior representation (see Fig.~\ref{fig:multihead}). Finally, we flattened the improved feature tensor $\mathbf{F''}$ to feed a linear layer producing an output dimension $\mathbf{h_{out}} = 2$ followed by a \textit{softmax} function for binary classification for protective and non-protective behavior.

\begin{figure*}[!htbp]
    \centering    {\includegraphics[width=0.9\linewidth]{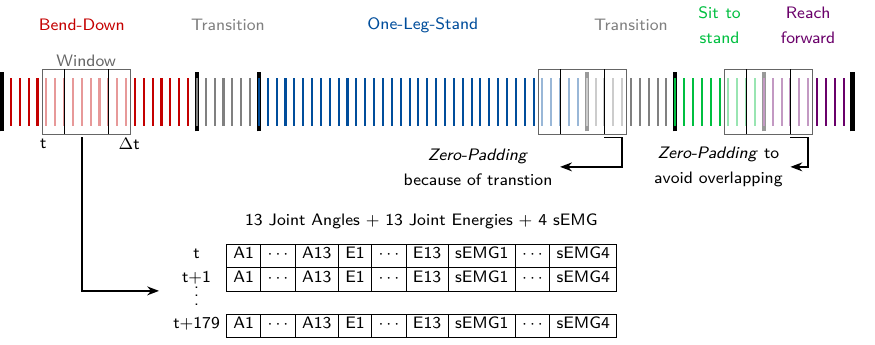}}
    \caption{The schematic of the sliding window segmentation technique for data preprocessing. Each window encapsulates a 3-second segment ($\Delta t=3$~s) with a 75\% overlap (hop size of 0.75~s). Within each window, 30 parameters are extracted per time step, encompassing 13 Joint Angles, 13 Joint Energies, and 4 sEMG values. Given the 60~Hz sampling rate, this translates to 180 distinct 30-dimensional vectors extracted per window. Zero-padding is applied if the 3-second window partially overlaps with a transition segment to ensure homogeneity within each window and avoid capturing transitions between activities. In this figure, we denote the transition activities as \com{Transition} segments.} \label{fig:window}
\end{figure*}

\section{Experiments and results}
\label{sec:experiments_results}

\subsection{EmoPain Dataset}
The EmoPain dataset~\cite{7173007} contains sensor recordings from physical exercises performed by 25 healthy individuals and 22 CLBP patients, selected to mimic everyday movements such as \com{reaching forward}, \com{bending down}, \com{sit-to-stand}, \com{stand-to-sit}, and \com{one-leg-stand}. These activities were executed at two difficulty levels, normal and difficult, with additional weights of 1~Kg and 2~Kg used for the \com{reaching forward} and \com{bending down} exercises to increase difficulty. For the \com{one-leg-stand}, participants used both their preferred and non-preferred legs. Specific instructions were provided for \com{sit-to-stand} and \com{stand-to-sit} exercises but not for the others.

Data were recorded using 18 wearable inertial measurement units (IMUs) for motion capture and 4 sEMG sensors for muscle activation. This study focuses on the inner \textbf{Joint Angle} and \textbf{Joint Energy} derived from IMU data, along with sEMG data from sensors placed on the lower and upper back. Protective behaviors were labeled in the complete sequence recordings by four medical professionals, including two physiotherapists and two psychologists, who segmented the videos and identified instances of protective behaviors or pain presence, covering seven distinct types of protective behaviors.

\subsection{Data preparation}
Our data preprocessing strategy was meticulously designed to accommodate the pre-filtered nature of the EmoPain database, focusing on normalization and segmentation to maintain data integrity and fully leverage the end-to-end model’s capabilities. Initially, Z-score normalization was applied to standardize the data across subjects, addressing variance discrepancies and ensuring uniformity in model input. Following normalization, we employed a sliding window approach, with a 75\% overlap for segmenting the continuous MoCAP and sEMG data streams. This method segments the data into 3-second windows, equivalent to 180 timesteps at a 60~Hz sampling rate (see Fig.~\ref{fig:window}). Such segmentation facilitates a spatiotemporal representation of size $180\times30$, encompassing 13 Joint Angles, 13 Joint Energies, and 4 sEMG values at each timestep. To augment the dataset and strengthen the model’s robustness against overfitting, we implemented random cropping with probabilities $p \in [0.05,0.15]$ and jittering by adding Gaussian noise with standard deviations $\sigma \in [0.05,0.15]$. Additionally, zero padding was used to ensure uniform window sizes across the dataset, particularly in scenarios where the sliding window could not capture 180 timesteps of data or when exercises overlapped. This comprehensive preprocessing approach, which follows that of Wang et al.~\cite{Wang_2021}, prepares the data and enhances the overall consistency of the dataset for the subsequent stages of model training and evaluation.

\subsection{Evaluation Metrics}
In line with established practices in analyzing the EmoPain dataset~\cite{PLAAN,emopainmultimodalmultilevel,emopainmultilevelcontext}, we employ a multi-metric evaluation strategy to ensure a robust and comprehensive assessment of our model's performance, underlining its effectiveness in addressing the challenges posed by the EmoPain dataset. Specifically, we use the mean $\mathrm{F1}$ score and the Matthews Correlation Coefficient (MCC) as primary metrics due to their effectiveness in handling class imbalances prevalent in this domain.

The mean $\mathrm{F1}$ score, represented as $\mathrm{Fm}$, harmonizes the balance between precision and recall, providing a holistic view of model accuracy across both positive and negative classes. It is computed using Eq.~\ref{eq:fm}.
\begin{align}
  \label{eq:fm}
  \mathrm{Precision} &= \frac{TP}{TP + FP}, \nonumber\\
  \mathrm{Recall} &= \frac{TP}{TP + FN}, \nonumber\\
  \mathrm{Fm} &= 2 \cdot \frac{\mathrm{Precision} \cdot \mathrm{Recall}}{\mathrm{Precision} + \mathrm{Recall}}
\end{align}

Moreover, the Matthews Correlation Coefficient (MCC), as defined in Eq.~\ref{eq:mcc}, offers a comprehensive measure that accounts for True and False Positives and Negatives (i.e., $TP$, $FP$, $TN$, and $FN$), making it particularly suitable for datasets with pronounced class imbalances.
\begin{equation}
  \label{eq:mcc}
  \resizebox{.9\columnwidth}{!}{$
      \mathrm{MCC} = \frac{TP \cdot TN - FP \cdot FN}{\sqrt{(TP+FP)(TP+FN)(TN+FP)(TN+FN)}}
  $}
\end{equation}

Both $\mathrm{Fm}$ and MCC require the selection of a specific threshold, which defines the operating point for these metrics. To mitigate the potential bias introduced by this selection and provide an evaluation that remains independent of operational choices, we also incorporate the Area Under the Precision-Recall Curve (AUC-PR) into our assessment framework. To assess the model's generalizability to unseen data and maximize the utilization of the available dataset, we employed a~\com{Leave-One-Subject-Out} (LOSO) cross-validation strategy~\cite{loso}. This approach uses all subjects except one for training and reserves the remaining subject for testing. This process is repeated, ensuring that each subject is evaluated once on unseen data.

\subsection{Implementation details}
The experiments were conducted on a computer with an Intel Core i7 11700 CPU, 32GB of RAM, and a GeForce RTX3060 GPU. The PyTorch version 1.13.1 served as the deep learning framework. The proposed network was trained using the Adam optimizer~\cite{kingma2017adam} due to its efficient handling of sparse and noisy gradients with a fixed learning rate of $\eta=0.001$ and a mini-batch size 40. Table~\ref{tab:implementation} summarizes the architecture details.

\begin{table}[!htbp]
\centering
\caption{L-SFAN Architecture Details}
\label{tab:implementation}
\resizebox{\columnwidth}{!}{%
\begin{tabular}{lllll}
\hline
\textbf{Layer Type}       & \textbf{Output Shape}    & \textbf{Kernel Size} & \textbf{Stride} & \textbf{Activation} \\ \hline
Input                     & $180 \times 30$          & -                    & -               & -                   \\
Conv Block 1              & $4 \times 180 \times 30$ & $3 \times 3$         & $1 \times 1$    & ReLU                \\
Max Pooling               & $4 \times 86 \times 30$ & $2 \times 1$         & $2 \times 1$    & -                   \\
Conv Block 2              & $16 \times 86 \times 30$ & $3 \times 3$         & $1 \times 1$    & ReLU                \\
Max Pooling               & $16 \times 42 \times 30$ & $2 \times 1$         & $2 \times 1$    & -                   \\
TAP                       & $16 \times 30$           & -                    & -               & -                   \\
Multi-Head Self-Attention & $5 \times 16 \times 30$  & -                    & -               & -                   \\
Feed-forward layer        & $16 \times 30$           & -                    & -               & -                   \\
Flatten                   & 480                      & -                    & -               & -                   \\
Linear                    & 2                        & -                    & -               & Softmax             \\ \hline
\end{tabular}%
}
\end{table}

\subsection{Experimental results}

In this section, we commence with an ablation study to evaluate the contributions of the key components of our proposed model. Firstly, we demonstrate the substantial performance improvement resulting from the use of temporal average pooling, highlighting the significance of spatial patterns in this context. Subsequently, we explore the contribution of the multi-head attention module. Finally, we thoroughly compare our proposed L-SFAN architecture and established state-of-the-art methods for protective behavior detection from multivariate time series data. This comparison serves to underscore the advantages of our approach and illustrate its potential impact.

\subsubsection{Relevance of the Spatial and Temporal Patterns}
\label{sec:pattern_relevance}

\begin{figure*}[!htpb]
    \centering
    \includegraphics[width=0.455\linewidth]{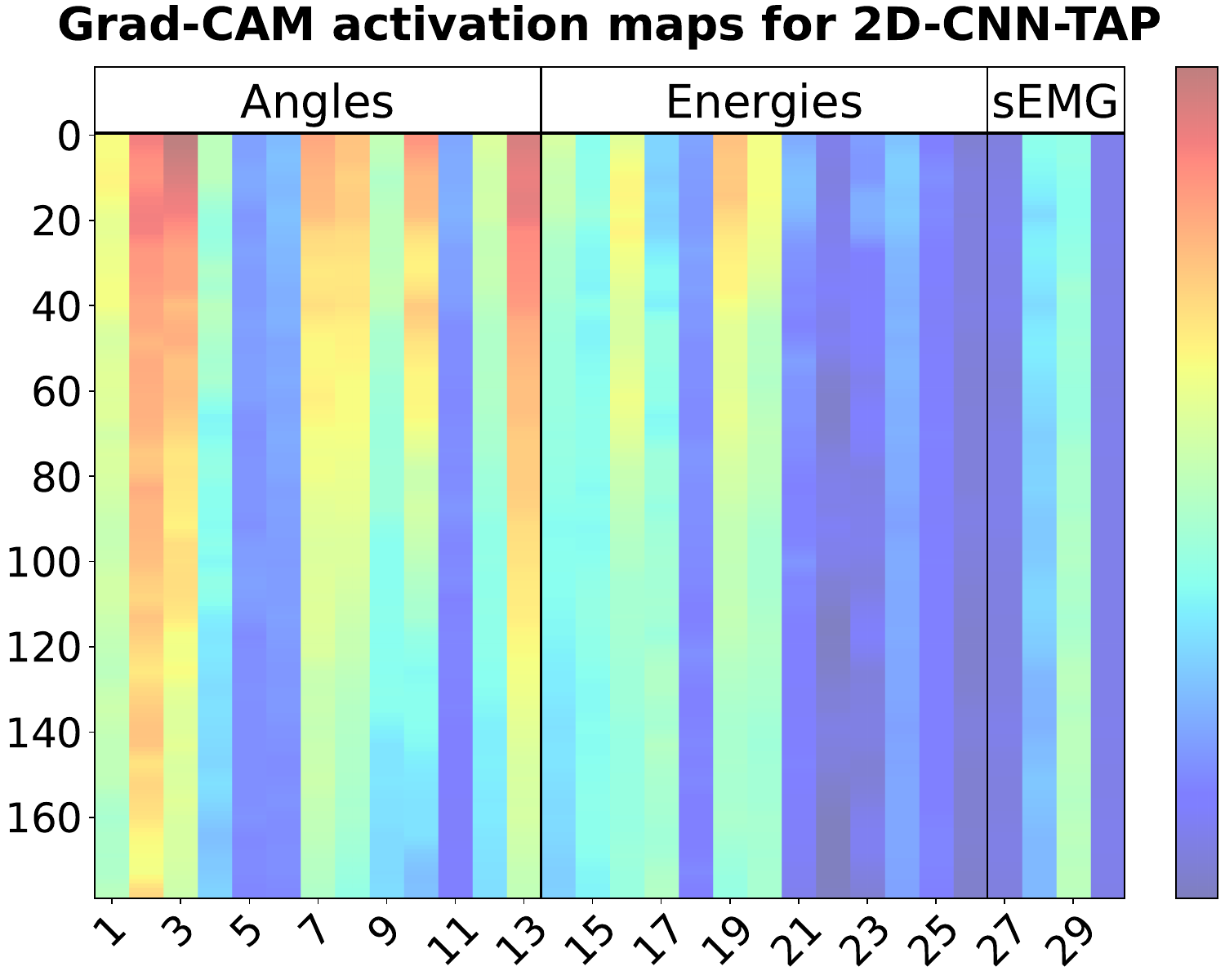}
    \hspace{1.5em}
    \includegraphics[width=0.455\linewidth]{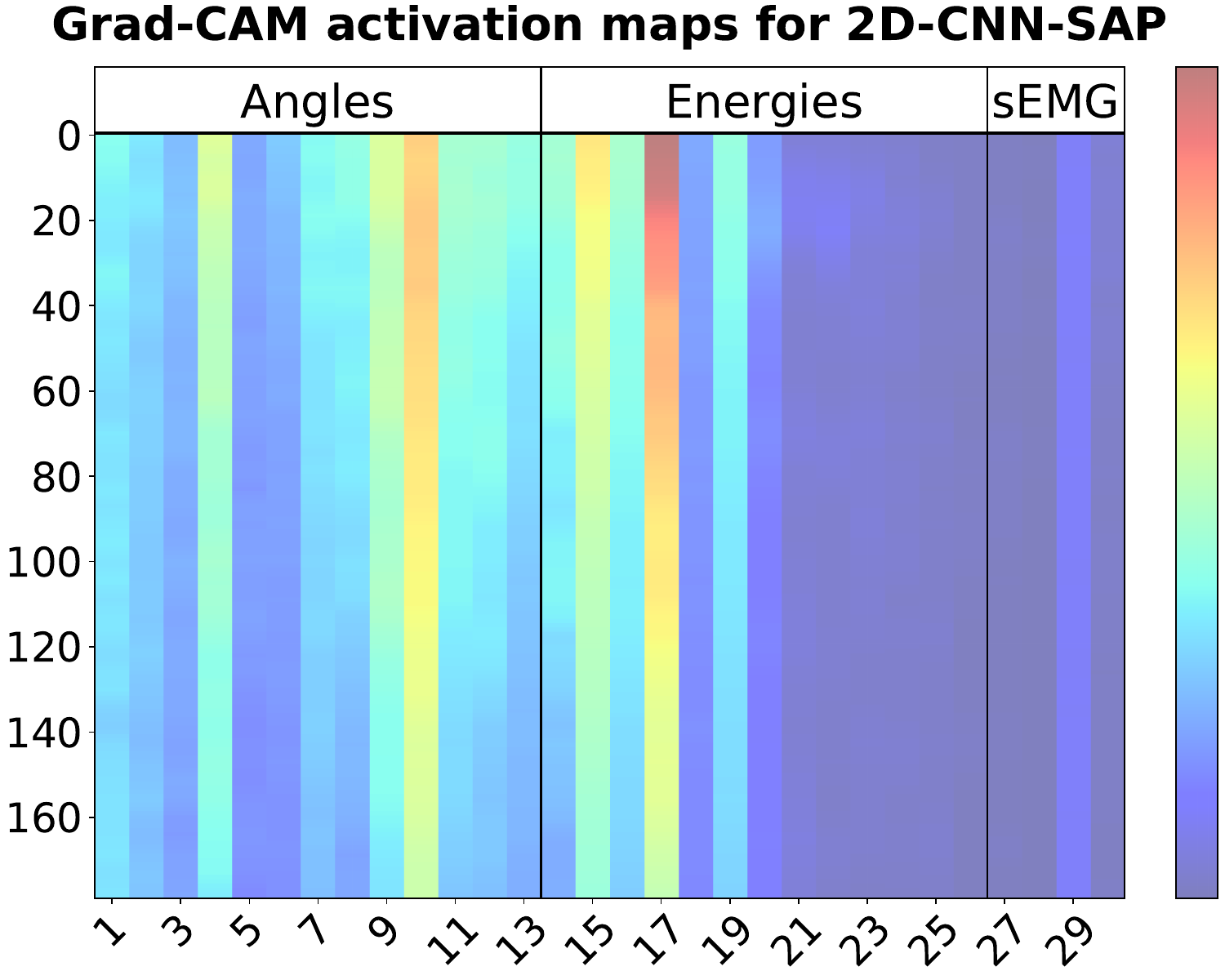}
    \caption{The heatmap depicts the activation levels across the input data ($180 \times 30$), with varying intensities indicating the significance of each input element towards the model's output. The graph on the left shows the activations for 2D-CNN-TAP, while the one on the right corresponds to 2D-CNN-SAP.}
    \label{fig:gradcam_TAP_SAP}
\end{figure*}

We conducted an ablation study comparing the feature extractor module's performance when focusing on temporal, spatial, and spatio-temporal patterns to determine the most effective information integration approach for protective behavior detection. This comparative exploration aimed to identify which approach best captures relevant patterns for PBD. Table~\ref{tab:integrationtype} summarizes the performance comparison of the 2D CNN architecture with different pooling strategies. The other aggregation methods included in this analysis are described next.

\begin{itemize}
  \item \textbf{GAP} (Global Average Pooling): This method averages the feature map $\mathbf{F}$ across both the temporal and spatial dimensions, resulting in a single scalar vector $\mathbf{F'} \in \mathbb{R}^{16}$.
  \item \textbf{SAP} (Spatial Average Pooling): Here, $\mathbf{F}$ is averaged only along the spatial dimension, leading to a new feature map $\mathbf{F'} \in \mathbb{R}^{16\times42}$ that retains the temporal resolution.
  \item \textbf{TAP} (Temporal Average Pooling): This method averages $\mathbf{F}$ across the temporal dimension, producing $\mathbf{F'} \in \mathbb{R}^{16\times30}$ that preserves the spatial resolution.
  \item \textbf{STAP} (Spatio-Temporal Average Pooling): This approach combines the features obtained from both TAP and SAP. The resulting tensor is a concatenation of $\mathbf{F'}_{TAP} \in \mathbb{R}^{16\times30}$ and $\mathbf{F'}_{SAP} \in \mathbb{R}^{16\times42}$, resulting in $\mathbf{F'} \in \mathbb{R}^{16\times30, 16\times42}$.
\end{itemize}

\begin{table}[!htbp]
  {\caption{Performance Comparison of Proposed CNN with Different Average Pooling Strategies.} \label{tab:integrationtype}} 
  \centering
    \begin{tabular}{lccc}
        \hline
        Feature Extractor & AUC & Fm & MCC \\
        \hline 
        2D-CNN & 0.676 & 0.673 & 0.257 \\
        2D-CNN-GAP & 0.645 & 0.673 & 0.169 \\
        2D-CNN-SAP & 0.699 & 0.688 & 0.299 \\
        2D-CNN-TAP & \textbf{0.819} & \textbf{0.752} & \textbf{0.481} \\
        2D-CNN-STAP & 0.786 & 0.725 & 0.414 \\
        \hline
      \end{tabular}
\end{table}

As shown in Table~\ref{tab:integrationtype}, emphasizing spatial information through temporal average pooling (2D-CNN-TAP) resulted in the highest performance, achieving an Area Under the Curve (AUC) of 0.819, F-measure of 0.752, and Matthews Correlation Coefficient (MCC) of 0.481. This observation aligns with the findings by Wang et al.~\cite{wang2019learning}, where incorporating a spatial attention module yielded superior results compared to a temporal attention module in their work.

Combining both spatial and temporal average pooling (2D-CNN-STAP) achieved the second-best performance. We hypothesize that this result primarily stems from including spatial information (2D-CNN-TAP) rather than their combined effect. This hypothesis is supported by the comparatively lower performance obtained when solely focusing on temporal patterns through spatial average pooling (2D-CNN-SAP).
 
Furthermore, the Grad-CAM visualization depicted in Fig.~\ref{fig:gradcam_TAP_SAP} illustrates the significance of the input data elements towards the model's output for a specific patient selected for illustration purposes. This visualization confirms the improved representation capabilities of 2D-CNN-TAP compared to 2D-CNN-SAP, as evidenced by the enhanced significance of the input data elements, which is quite compelling. 

\subsubsection{Multi-head Attention Module}
\label{sec:multi-head-attention}

To assess the contribution of the multi-head self-attention module, we incorporated it into the model using the best-performing feature extractor (i.e., \mbox{2D-CNN-TAP}). The performance contribution of this subsystem is shown in Table~\ref{tab:multi-head-attention}. As observed, the multi-head attention module led to a significant improvement in performance metrics. 

\begin{figure}[!htpb]
    \centering
    {\includegraphics[width=\linewidth]{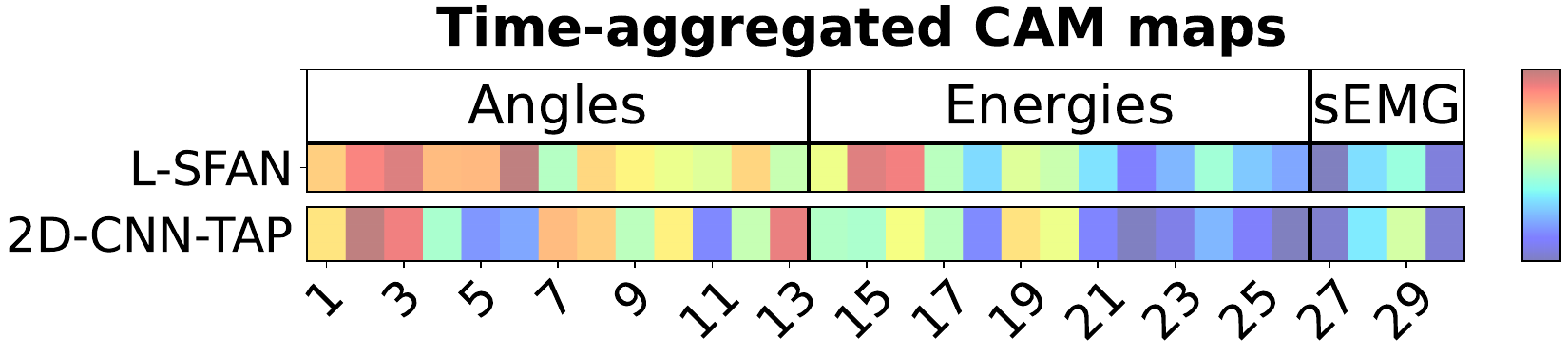}}
    \caption{Comparison of activation maps between L-SFAN (top) and 2D-CNN-TAP (bottom) architectures. The 2D maps were aggregated over time and normalized to highlight spatial patterns.}
    \label{fig:gradcam_TAP_ATT}
\end{figure}

In addition, Fig.~\ref{fig:gradcam_TAP_ATT} demonstrates the Grad-CAM visualization for both architectures. In this case, we have used a time-aggregated version to focus solely on spatial patterns for the same patient. One can observe that the visualization is compelling, demonstrating how including the attention mechanism results in higher activation levels.

\begin{table}[!htbp]
  \caption{Performance comparison of 2D-CNN-TAP vs. L-SFAN, demonstrating the impact of the multi-head attention module. } 
  \label{tab:multi-head-attention}
  \centering
    \begin{tabular}{lcccc}
        \hline
        Architecture & AUC & Fm & MCC & \#Parameters \\
        \hline 
        2D-CNN-TAP & 0.819 & 0.752 & 0.481 & 2,582 \\
        L-SFAN &\textbf{0.849} & \textbf{0772} & \textbf{0.51} & 8,282\\ \hline
    \end{tabular}
\end{table}

However, it is important to acknowledge the trade-off involved. While the multi-head attention module demonstrably enhances performance, it comes at the cost of a nearly threefold increase in trainable parameters, as shown in Table~\ref{tab:comparison}. This highlights the importance of considering both performance gains and computational complexity when selecting model components.

\subsubsection{Comparison with the state-of-the-art}

Table~\ref{tab:comparison} presents a comprehensive comparison of the proposed architecture, \mbox{L-SFAN}, with various state-of-the-art methods on the EmoPain dataset. The evaluation metrics used are Area Under the Curve (AUC), mean F1 score (Fm), Matthews Correlation Coefficient (MCC), and the number of trainable parameters.

\begin{table}[!htbp]
  {\caption{Performance comparison of the proposed model with state-of-the-art methods on the EmoPain dataset.} 
  \label{tab:comparison}}
  \centering
    \resizebox{\columnwidth}{!}{%
    \begin{tabular}{lcccc}
        \hline
        Methods & AUC & Fm & MCC & \#Parameters \\
        \hline
        \multicolumn{5}{l}{\cellcolor[HTML]{EFEFEF}State-of-the-art methods} \\ \hline
        Stacked-LSTM~\cite{lstm2}  & 0.622 & 0.662 & 0.192 & 25,154 \\
        Dual-Stream LSTM~\cite{Wang_2021} & 0.618 & 0.671 & 0.238 & 16,258 \\
        BANet~\cite{wang2019learning} & 0.663 & 0.668 & 0.247 & 2,131 \\
        MiMT~\cite{mimt} & 0.648 & 0.722 & 0.327 & 1,038 \\
        LSTM+GCN~\cite{wang2021leveraging} & 0.690 & 0.731 & 0.365 & 93,414 \\
        Sparse VAE~\cite{sparsevae} & 0.641 & 0.658 & 0.312 & 64,572\\
        Gaussian VAE~\cite{gaussianvae} & 0.622 & 0.631 & 0.185 & 111,826\\
        GRU RNN~\cite{moha} & \textbf{0.855} & 0.765 & 0.411 & 140,352\\
        \hline
        \multicolumn{5}{l}{\cellcolor[HTML]{EFEFEF}Proposed methods} \\ \hline
        2D-CNN-TAP & 0.819 & 0.752 & 0.481 & 2,582\\
        L-SFAN & 0.849 & \textbf{0.772} & \textbf{0.51} & 8,282\\
        \hline
      \end{tabular}%
    }
\end{table}

\begin{figure}[!htpb]
    \centering
    {\includegraphics[width=0.95\linewidth, scale=0.95]{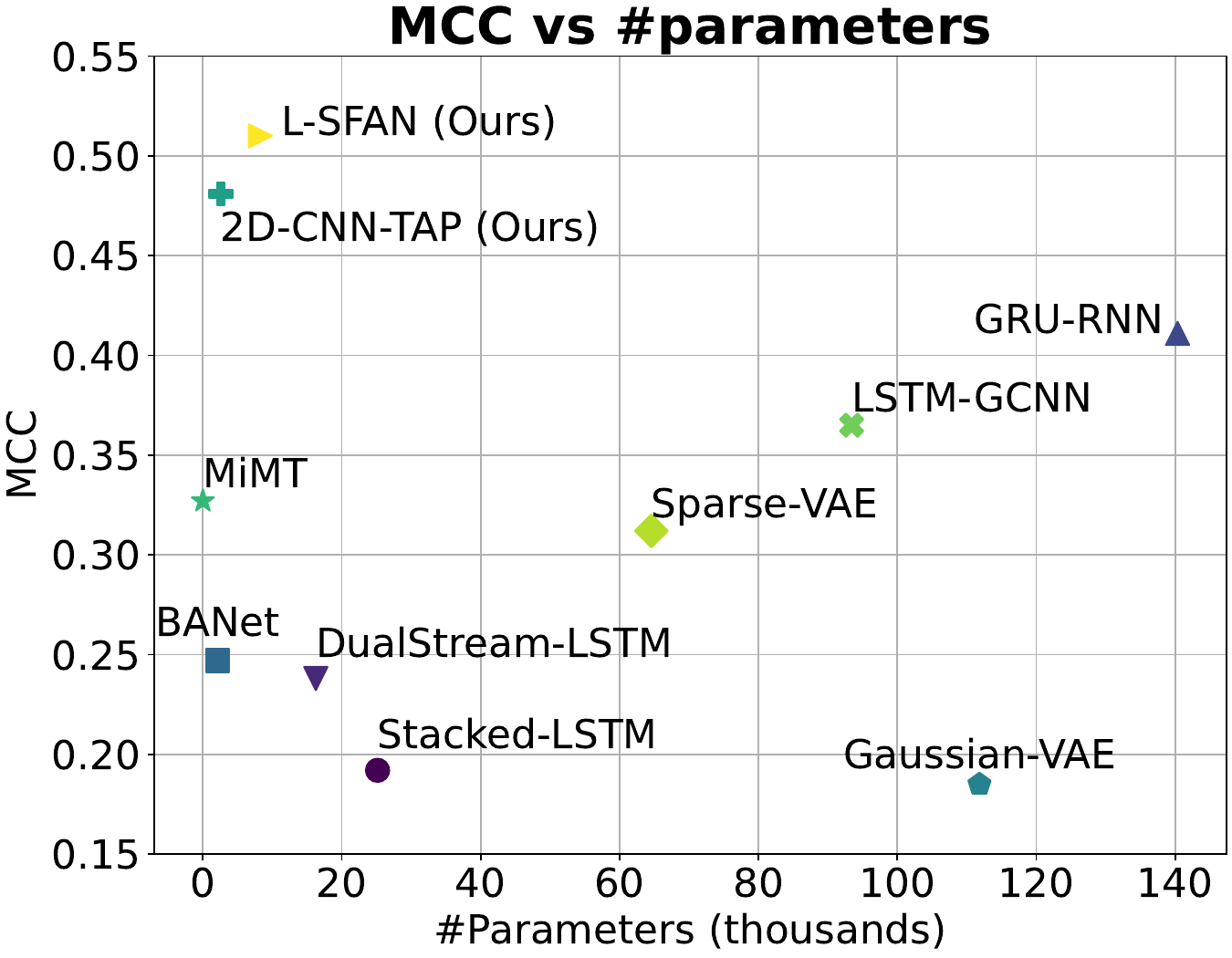}}
    \caption{Performance comparison of various state-of-the-art models in terms of MCC versus the number of trainable parameters.}
    \label{fig:scatter}
\end{figure}

As shown in Table~\ref{tab:comparison}, the proposed L-SFAN architecture achieves the highest MCC (0.510) and Fm (0.772) among all compared methods. While the proposed model in~\cite{moha} achieves slightly higher AUC (0.855), it comes at the expense of a significantly higher number of parameters (140,352 compared to 8,282 for L-SFAN). This highlights the efficiency and effectiveness of the proposed architecture, particularly in the context of a limited dataset where overfitting can be a significant concern. We attribute the superior performance of L-SFAN to two key factors:
\begin{enumerate}
    \item \textbf{Effective Spatial Pattern Extraction:} The emphasis on spatial patterns through temporal average pooling (TAP), along with the synergistic integration of CNNs and self-attention, enables the model to efficiently capture the spatial relationships crucial to our task. To further illustrate the efficacy of the proposed model in preserving crucial spatial information for protective behavior detection, we used Grad-CAM~\cite{Selvaraju_2017_ICCV} to visualize the activation levels across the 13 joint angles, 13 joint energies, and 4 sEMG channels averaged over the temporal dimension (see Fig.~\ref{fig:gradcam_TAP_SAP}). As previously discussed, this analysis highlighted how the data representation generated by L-SFAN enhances the significance of the input data elements for the task, making it more effective.
    \item \textbf{Reduced Model Complexity:} The lightweight design of the L-SFAN architecture, with significantly fewer trainable parameters compared to most state-of-the-art methods (except BANet~\cite{wang2019learning} and MiMT~\cite{mimt}), substantially mitigates the risk of overfitting, especially when dealing with datasets like EmoPain that have a limited number of samples (see Fig.~\ref{fig:scatter}).
\end{enumerate}

\section{Conclusion and Future Research Directions}
\label{sec:coclusion}
This study addressed the challenging task of automatically detecting protective behaviors from multivariate time series data using a novel lightweight CNN architecture with attention. The proposed architecture, named L-SFAN, leverages a 2D CNN-based feature extraction backbone specifically focused on spatial patterns, followed by a multi-head self-attention mechanism to refine the extracted features. L-SFAN demonstrates competitive performance compared to existing state-of-the-art methods, achieving an  Area Under the Precision-Recall Curve of 84.9\%, F1-measure of 86.8\%, and Matthews Correlation Coefficient of 0.744 on the EmoPain dataset.

Our comparative analysis reveals the advantages of the L-SFAN approach, particularly its ability to effectively uncover key spatial patterns for the task from multivariate sensor data using a lightweight architecture. Our ablation study further provides valuable insights into the model's subsystems, highlighting the crucial role of spatial patterns and self-attention module in achieving optimal performance.

Moreover, the lightweight nature of the proposed architecture deserves appropriate attention, as it represents a significant step towards the development of advanced and scalable AI-based tools for resource-constrained settings such as home environments. This is particularly crucial when access to effective rehabilitation in traditional medical settings is significantly limited.

However, limitations exist that offer opportunities for future research. Firstly, while L-SFAN demonstrates promising results, further exploration of advanced self-attention mechanisms could improve performance (maintaining a lightweight architecture). Additionally, the study focused on a specific type of protective behavior using a single dataset. Expanding the analysis to diverse datasets and incorporating broader variations of protective behavior definitions would enhance the model's generalization ability and robustness.  Finally, to advance towards affordable at-home monitoring, addressing the selection of the most relevant sensors is crucial.

\section*{References}

\bibliographystyle{IEEEtran}
\bibliography{bibliography}

\begin{thebibliography}{10}
\providecommand{\url}[1]{#1}
\csname url@samestyle\endcsname
\providecommand{\newblock}{\relax}
\providecommand{\bibinfo}[2]{#2}
\providecommand{\BIBentrySTDinterwordspacing}{\spaceskip=0pt\relax}
\providecommand{\BIBentryALTinterwordstretchfactor}{4}
\providecommand{\BIBentryALTinterwordspacing}{\spaceskip=\fontdimen2\font plus
\BIBentryALTinterwordstretchfactor\fontdimen3\font minus \fontdimen4\font\relax}
\providecommand{\BIBforeignlanguage}[2]{{%
\expandafter\ifx\csname l@#1\endcsname\relax
\typeout{** WARNING: IEEEtran.bst: No hyphenation pattern has been}%
\typeout{** loaded for the language `#1'. Using the pattern for}%
\typeout{** the default language instead.}%
\else
\language=\csname l@#1\endcsname
\fi
#2}}
\providecommand{\BIBdecl}{\relax}
\BIBdecl

\bibitem{Yong2021}
R.~J. Yong, P.~M. Mullins, and N.~Bhattacharyya, ``{Prevalence of chronic pain among adults in the United States},'' \emph{{PAIN}}, vol. 163, no.~2, pp. e328--e332, 2022.

\bibitem{ROY2022123}
R.~Roy, S.~Gal{\'a}n, E.~S{\'a}nchez-Rodr{\'\i}guez, M.~Racine, E.~Sol{\'e}, M.~P. Jensen, and J.~Mir{\'o}, ``{Cross-National Trends of Chronic Back Pain in Adolescents: Results From the HBSC Study, 2001-2014},'' \emph{{The Journal of Pain}}, vol.~23, no.~1, pp. 123--130, 2022.

\bibitem{Fayaz2016}
A.~Fayaz, P.~Croft, R.~M. Langford, L.~J. Donaldson, and G.~T. Jones, ``\BIBforeignlanguage{en}{Prevalence of chronic pain in the {UK}: a systematic review and meta-analysis of population studies},'' \emph{\BIBforeignlanguage{en}{BMJ Open}}, vol.~6, no.~6, p. e010364, Jun. 2016.

\bibitem{Aneesha}
A.~Singh, N.~Bianchi-Berthouze, and A.~C. Williams, ``Supporting everyday function in chronic pain using wearable technology,'' in \emph{Proceedings of the 2017 CHI Conference on Human Factors in Computing Systems}, ser. CHI'17.\hskip 1em plus 0.5em minus 0.4em\relax Association for Computing Machinery, 2017, pp. 3903–--3915.

\bibitem{Leeuw}
G.~van~der Leeuw, L.~H.~P. Eggermont, L.~Shi, W.~P. Milberg, A.~L. Gross, J.~M. Hausdorff, J.~F. Bean, and S.~G. Leveille, ``{Pain and Cognitive Function Among Older Adults Living in the Community},'' \emph{The Journals of Gerontology: Series A}, vol.~71, no.~3, pp. 398--405, 2015.

\bibitem{KEEFE1982363}
\BIBentryALTinterwordspacing
F.~J. Keefe and A.~R. Block, ``Development of an observation method for assessing pain behavior in chronic low back pain patients,'' \emph{Behavior Therapy}, vol.~13, no.~4, pp. 363--375, 1982. [Online]. Available: \url{https://www.sciencedirect.com/science/article/pii/S0005789482800014}
\BIBentrySTDinterwordspacing

\bibitem{VLAEYEN2000317}
\BIBentryALTinterwordspacing
J.~W. Vlaeyen and S.~J. Linton, ``Fear-avoidance and its consequences in chronic musculoskeletal pain: a state of the art,'' \emph{Pain}, vol.~85, no.~3, pp. 317--332, 2000. [Online]. Available: \url{https://www.sciencedirect.com/science/article/pii/S0304395999002420}
\BIBentrySTDinterwordspacing

\bibitem{ashtari2022multi}
M.~Ashtari-Majlan, A.~Seifi, and M.~M. Dehshibi, ``{A Multi-Stream Convolutional Neural Network for Classification of Progressive MCI in Alzheimer’s Disease Using Structural MRI Images},'' \emph{{IEEE Journal of Biomedical and Health Informatics}}, vol.~26, no.~8, pp. 3918--3926, 2022.

\bibitem{dehshibi2023deep}
M.~M. Dehshibi, B.~Baiani, G.~Pons, and D.~Masip, ``{A Deep Multimodal Learning Approach to Perceive Basic Needs of Humans From Instagram Profile},'' \emph{{IEEE Transactions on Affective Computing}}, vol.~14, no.~2, pp. 944--956, 2023.

\bibitem{lstm2}
C.~Wang, T.~A. Olugbade, A.~Mathur, A.~C. De~C.~Williams, N.~D. Lane, and N.~Bianchi-Berthouze, ``Recurrent network based automatic detection of chronic pain protective behavior using mocap and semg data,'' in \emph{Proceedings of the ACM International Symposium on Wearable Computers}.\hskip 1em plus 0.5em minus 0.4em\relax Association for Computing Machinery, 2019, pp. 225–--230.

\bibitem{moha}
M.~M. Dehshibi, T.~Olugbade, F.~Diaz-de Maria, N.~Bianchi-Berthouze, and A.~Tajadura-Jimenez, ``{Pain Level and Pain-Related Behaviour Classification Using GRU-Based Sparsely-Connected RNNs},'' \emph{{IEEE Journal of Selected Topics in Signal Processing}}, vol.~17, no.~3, pp. 677--688, 2023.

\bibitem{wang2021leveraging}
C.~Wang, Y.~Gao, A.~Mathur, A.~C. D.~C. Williams, N.~D. Lane, and N.~Bianchi-Berthouze, ``Leveraging activity recognition to enable protective behavior detection in continuous data,'' 2021.

\bibitem{wang2019learning}
C.~Wang, M.~Peng, T.~A. Olugbade, N.~D. Lane, A.~C. D.~C. Williams, and N.~Bianchi-Berthouze, ``Learning bodily and temporal attention in protective movement behavior detection,'' 2019.

\bibitem{Martineau2020}
J.~Casta{\~{n}}o, S.~Mart{\'i}nez-Fern{\'a}ndez, X.~Franch, and J.~Bogner, ``{Exploring the Carbon Footprint of Hugging Face's ML Models: A Repository Mining Study},'' in \emph{{ACM/IEEE International Symposium on Empirical Software Engineering and Measurement (ESEM)}}, 2023, pp. 1--12.

\bibitem{emopainmultimodalmultilevel}
M.~T. Uddin and S.~Canavan, ``Multimodal multilevel fusion for sequential protective behavior detection and pain estimation,'' in \emph{2020 15th IEEE International Conference on Automatic Face and Gesture Recognition (FG 2020)}, 2020, pp. 844--848.

\bibitem{emopainmultilevelcontext}
K.~N. Phan, N.~K. Iyortsuun, S.~Pant, H.-J. Yang, and S.-H. Kim, ``Pain recognition with physiological signals using multi-level context information,'' \emph{IEEE Access}, vol.~11, pp. 20\,114--20\,127, 2023.

\bibitem{Selvaraju_2017_ICCV}
R.~R. Selvaraju, M.~Cogswell, A.~Das, R.~Vedantam, D.~Parikh, and D.~Batra, ``{Grad-CAM: Visual Explanations from Deep Networks via Gradient-Based Localization},'' in \emph{{2017 IEEE International Conference on Computer Vision (ICCV)}}.\hskip 1em plus 0.5em minus 0.4em\relax IEEE, 2017, pp. 618--626.

\bibitem{7173007}
M.~S.~H. Aung, S.~Kaltwang, B.~Romera-Paredes, B.~Martinez, A.~Singh, M.~Cella, M.~Valstar, H.~Meng, A.~Kemp, M.~Shafizadeh, A.~C. Elkins, N.~Kanakam, A.~de~Rothschild, N.~Tyler, P.~J. Watson, A.~C. d.~C. Williams, M.~Pantic, and N.~Bianchi-Berthouze, ``{The Automatic Detection of Chronic Pain-Related Expression: Requirements, Challenges and the Multimodal EmoPain Dataset},'' \emph{{IEEE Transactions on Affective Computing}}, vol.~7, no.~4, pp. 435--451, 2016.

\bibitem{Wang_2021}
\BIBentryALTinterwordspacing
C.~Wang, T.~A. Olugbade, A.~Mathur, A.~C. D.~C. Williams, N.~D. Lane, and N.~Bianchi-Berthouze, ``Chronic pain protective behavior detection with deep learning,'' \emph{{ACM} Transactions on Computing for Healthcare}, vol.~2, no.~3, pp. 1--24, 2021. [Online]. Available: \url{https://doi.org/10.1145\%2F3449068}
\BIBentrySTDinterwordspacing

\bibitem{PLAAN}
Y.~Li, S.~Ghosh, and J.~Joshi, ``Plaan: Pain level assessment with anomaly-detection based network,'' \emph{Journal on Multimodal User Interfaces}, vol.~15, no.~4, pp. 359--372, 2021.

\bibitem{bahdanau2016neural}
D.~Bahdanau, K.~Cho, and Y.~Bengio, ``Neural machine translation by jointly learning to align and translate,'' 2016.

\bibitem{10182360}
M.~T. Uddin, G.~Zamzmi, and S.~Canavan, ``{Cooperative Learning for Personalized Context-Aware Pain Assessment From Wearable Data},'' \emph{{IEEE Journal of Biomedical and Health Informatics}}, vol.~27, no.~11, pp. 5260--5271, 2023.

\bibitem{alexnet}
\BIBentryALTinterwordspacing
A.~Krizhevsky, I.~Sutskever, and G.~E. Hinton, ``Imagenet classification with deep convolutional neural networks,'' in \emph{Advances in Neural Information Processing Systems 25}, F.~Pereira, C.~J.~C. Burges, L.~Bottou, and K.~Q. Weinberger, Eds.\hskip 1em plus 0.5em minus 0.4em\relax Curran Associates, Inc., 2012, pp. 1097--1105. [Online]. Available: \url{http://papers.nips.cc/paper/4824-imagenet-classification-with-deep-convolutional-neural-networks.pdf}
\BIBentrySTDinterwordspacing

\bibitem{long2014fully}
\BIBentryALTinterwordspacing
J.~Long, E.~Shelhamer, and T.~Darrell, ``Fully convolutional networks for semantic segmentation,'' 2014, cite arxiv:1411.4038Comment: to appear in CVPR (2015). [Online]. Available: \url{http://arxiv.org/abs/1411.4038}
\BIBentrySTDinterwordspacing

\bibitem{dehshibi2024beenet}
\BIBentryALTinterwordspacing
M.~M. Dehshibi and D.~Masip, ``{BEE-NET: A deep neural network to identify in-the-wild Bodily Expression of Emotions},'' pp. 1--10, 2024. [Online]. Available: \url{https://arxiv.org/abs/2402.13955}
\BIBentrySTDinterwordspacing

\bibitem{dehshibi2023ADVISE}
M.~M. Dehshibi, M.~Ashtari-Majlan, G.~Adhane, and D.~Masip, ``{ADVISE: ADaptive feature relevance and VISual Explanations for convolutional neural networks},'' \emph{{The Visual Computer}}, pp. 1--13, 2023.

\bibitem{ashtarimajlan2023deep}
\BIBentryALTinterwordspacing
M.~Ashtari-Majlan, M.~M. Dehshibi, and D.~Masip, ``{Deep Learning and Computer Vision for Glaucoma Detection: A Review},'' pp. 1--20, 2023. [Online]. Available: \url{https://arxiv.org/abs/2307.16528}
\BIBentrySTDinterwordspacing

\bibitem{7870510}
B.~Zhao, H.~Lu, S.~Chen, J.~Liu, and D.~Wu, ``Convolutional neural networks for time series classification,'' \emph{Journal of Systems Engineering and Electronics}, vol.~28, no.~1, pp. 162--169, 2017.

\bibitem{loffe2015}
\BIBentryALTinterwordspacing
S.~Ioffe and C.~Szegedy, ``Batch normalization: Accelerating deep network training by reducing internal covariate shift,'' in \emph{Proceedings of the 32nd International Conference on Machine Learning}, ser. Proceedings of Machine Learning Research, F.~Bach and D.~Blei, Eds., vol.~37.\hskip 1em plus 0.5em minus 0.4em\relax PMLR, 2015, pp. 448--456. [Online]. Available: \url{https://proceedings.mlr.press/v37/ioffe15.html}
\BIBentrySTDinterwordspacing

\bibitem{9310460}
H.~Wang and M.~Tu, ``Enhancing attention models via multi-head collaboration,'' in \emph{2020 International Conference on Asian Language Processing (IALP)}.\hskip 1em plus 0.5em minus 0.4em\relax IEEE, 2020, pp. 19--23.

\bibitem{LI202114}
J.~Li, X.~Wang, Z.~Tu, and M.~R. Lyu, ``On the diversity of multi-head attention,'' \emph{Neurocomputing}, vol. 454, pp. 14--24, 2021.

\bibitem{loso}
T.-T. Wong, ``Performance evaluation of classification algorithms by k-fold and leave-one-out cross validation,'' \emph{{Pattern Recognition}}, vol.~48, no.~9, pp. 2839--2846, 2015.

\bibitem{kingma2017adam}
\BIBentryALTinterwordspacing
D.~P. Kingma and J.~Ba, ``{Adam: A Method for Stochastic Optimization},'' in \emph{3rd International Conference on Learning Representations, ICLR 2015}, 2015, pp. 1--15. [Online]. Available: \url{http://arxiv.org/abs/1412.6980}
\BIBentrySTDinterwordspacing

\bibitem{mimt}
T.~Olugbade, N.~Gold, A.~C. d.~C. Williams, and N.~Bianchi-Berthouze, ``{A Movement in Multiple Time Neural Network for Automatic Detection of Pain Behaviour},'' in \emph{International Conference on Multimodal Interaction}.\hskip 1em plus 0.5em minus 0.4em\relax Association for Computing Machinery, 2021, pp. 442–--445.

\bibitem{sparsevae}
\BIBentryALTinterwordspacing
L.~Antelmi, N.~Ayache, P.~Robert, and M.~Lorenzi, ``{Sparse Multi-Channel Variational Autoencoder for the Joint Analysis of Heterogeneous Data},'' in \emph{{Proceedings of the 36th International Conference on Machine Learning}}.\hskip 1em plus 0.5em minus 0.4em\relax {Proceedings of Machine Learning Research}, 2019, pp. 302--311. [Online]. Available: \url{https://proceedings.mlr.press/v97/antelmi19a.html}
\BIBentrySTDinterwordspacing

\bibitem{gaussianvae}
\BIBentryALTinterwordspacing
{Guo, Yifan and Liao, Weixian and Wang, Qianlong and Yu, Lixing and Ji, Tianxi and Li, Pan}, ``{Multidimensional Time Series Anomaly Detection: A GRU-based Gaussian Mixture Variational Autoencoder Approach},'' in \emph{{Proceedings of The 10th Asian Conference on Machine Learning}}.\hskip 1em plus 0.5em minus 0.4em\relax {Proceedings of Machine Learning Research}, 2018, pp. 97--112. [Online]. Available: \url{https://proceedings.mlr.press/v95/guo18a.html}
\BIBentrySTDinterwordspacing

\end{thebibliography}

\vfill
\end{document}